\documentclass[10pt,aps,prd,preprintnumbers,showpacs,nofootinbib,superscriptaddress,notitlepage]{revtex4-1}

\usepackage{graphicx}
\usepackage{color}
\usepackage{rotating}
\usepackage{amsmath}
\usepackage{amssymb}
\usepackage{enumerate}
\usepackage[colorlinks=true,citecolor=blue,urlcolor=blue]{hyperref}

\newcommand{\PRE}[1]{{#1}} 

\newcommand{\cm}{\ensuremath{\mathrm{cm}}}
\newcommand{\s}{\ensuremath{\mathrm{s}}}
\newcommand{\GeV}{\ensuremath{\mathrm{GeV}}}
\newcommand{\MeV}{\ensuremath{\mathrm{MeV}}}

\newcommand{\PhiPP}{\Phi_\textrm{PP}}
\newcommand{\Aeff}{A_\textrm{eff}}
\newcommand{\Abareff}{\bar{A}_\textrm{eff}}
\newcommand{\Nbar}{\overline{N}}

\newcommand{\DM}{\textrm{DM}}
\newcommand{\tot}{\textrm{tot}}
\newcommand{\bgd}{\textrm{bgd}}
\newcommand{\obs}{\textrm{obs}}
\newcommand{\bound}{\textrm{bound}}

\begin{document}


\title{Model-independent constraints on dark matter annihilation\\ in dwarf spheroidal galaxies}

\author{Kimberly K. Boddy}
\affiliation{\mbox{Department of Physics \& Astronomy,
Johns Hopkins University, Baltimore, MD 21218, USA}}
\affiliation{\mbox{Department of Physics \& Astronomy,
University of Hawai'i, Honolulu, HI 96822, USA}}

\author{Jason Kumar}
\affiliation{\mbox{Department of Physics \& Astronomy,
University of Hawai'i, Honolulu, HI 96822, USA}}

\author{Danny Marfatia}
\affiliation{\mbox{Department of Physics \& Astronomy,
University of Hawai'i, Honolulu, HI 96822, USA}}

\author{Pearl Sandick\PRE{\vspace*{.1in}}}
\affiliation{\mbox{Department of Physics and Astronomy,
University of Utah, Salt Lake City, UT 84112, USA}}

\begin{abstract}
\PRE{\vspace*{.1in}}

We present a general, model-independent formalism for determining bounds on the production of photons in dwarf spheroidal galaxies via dark matter annihilation, applicable to any set of assumptions about dark matter particle physics or astrophysics.
As an illustration, we analyze gamma-ray data from the Fermi Large Area Telescope to constrain a variety of nonstandard dark matter models, several of which have not previously been studied in the context of dwarf galaxy searches.

\end{abstract}

\maketitle


\section{Introduction}

Dwarf spheroidal galaxies (dSphs) are the cleanest environment within which to search for photons arising from dark matter (DM) annihilation.
As a result, a variety of analyses have focused on this set of targets.
In general, a search strategy tailored to a particular model will tend to provide better sensitivity to that model, but this added sensitivity comes at a cost: for each model, the analysis must be done from scratch.
Our goal in this work is to provide the most general, model-independent analysis of the latest gamma-ray data from the Fermi Large Area Telescope (LAT) for dwarf spheroidal galaxies, which can be easily applied to any choice of particle physics model and to any choice of astrophysical parameters.
We find that this level of generality can be achieved with only a modest loss of sensitivity.

Since 2008, the Fermi-LAT has been collecting gamma-ray data in the energy range of $20~\MeV$ to $300~\GeV$, ideal for searching for weakly-interacting massive particle DM annihilations.
The Fermi-LAT collaboration uses a likelihood analysis to fit the spatial and spectral features of dSphs to obtain upper limits on the annihilation cross section as a function of DM mass~\cite{Abdo:2010ex,Ackermann:2011wa,Ackermann:2013yva,Ackermann:2015zua,Fermi-LAT:2016uux}.
They model the Galactic and isotropic diffuse emission, account for point-like sources from the latest LAT source catalog, and incorporate uncertainties in the determination of astrophysical $J$-factors.
Instead of relying on background/foreground modeling, others have determined the diffuse emission empirically through frequentist~\cite{GeringerSameth:2011iw,GeringerSameth:2014qqa} and Bayesian~\cite{Mazziotta:2012ux} methods, which yield comparable results to the Fermi analyses.
The analysis in Ref.~\cite{GeringerSameth:2011iw} utilizes only the overall number of counts, discarding  spectral information, in order to make the analysis more generic.

The analysis presented here differs from prior analyses in several important ways.
We follow Refs.~\cite{GeringerSameth:2011iw,GeringerSameth:2014qqa} to obtain the background/foreground distribution (which for brevity, we refer to as the background distribution for the remainder of this paper) for each dSph, and we provide these distributions as a supplementary file~\cite{supp}.
The resulting background distribution, along with the number of photons observed by Fermi-LAT in the direction of the dSph, provides a statistical limit on the number of photons attributable to DM annihilation.
To increase statistical power, we stack the dSphs, weighting all photon events equally.
In contrast, Refs.~\cite{GeringerSameth:2011iw,GeringerSameth:2014qqa} perform an analysis that weights photons events by the astrophysical $J$-factor of the dSph of origin, the reconstructed energy, and the angular distance from the center of the dSph.
A simple stacked analysis has the advantage of separating the information contained within the Fermi-LAT data, the value of the $J$-factor, and the details of the DM annihilation.
We also use the latest Pass8 data set from Fermi-LAT, compared to the Pass7 and Pass7 Reprocessed data sets used in Ref.~\cite{GeringerSameth:2011iw} and Ref.~\cite{GeringerSameth:2014qqa}, respectively.
In addition to having more photon events, we also use the updated 4-year Fermi-LAT point source catalog (3FGL) to better reject contamination from known point sources in the background distributions.

Our general procedure provides portability for the particle physics community to obtain bounds on any model without having to rely on a particular set of $J$-factors (as is the case when using the flux upper limits in Ref.~\cite{Fermi-LAT:2016uux}, for example) and without having to run a Fermi analysis from scratch.
It can be easily extended to a broad variety of scenarios for which prior analyses are inapplicable.
For example, although there are a wide variety of analyses for two-body $s$-wave DM annihilation to Standard Model fermions ($XX \rightarrow \bar f f$), they cannot be applied to the case in which the dominant annihilation process in the current epoch is internal bremsstrahlung ($XX \rightarrow \bar f f \gamma$), which can result in a photon spectrum that is very different from the standard two-body case.
Similarly, current analyses that study multiple dSphs use particular assumptions for the DM density profile in each dSph, and thus cannot be simply generalized to different values of the $J$-factors.
By contrast, our analysis can be easily generalized not only to different choices of $J$-factors, but also to the case of velocity-dependent annihilation, in which case the effective $J$-factor depends on the DM velocity profile.

The plan of this work is as follows.
In Section~\ref{sec:analysis}, we describe the general analysis framework.
In Section~\ref{sec:results}, we use this framework to present bounds on several DM particle and astrophysics scenarios, including several scenarios for which no previous bounds have been exhibited.
We conclude with a discussion of our results in Section~\ref{sec:conclusions}.


\section{Analysis Framework}
\label{sec:analysis}

The expected number of photons within a given energy range that arise from DM annihilation in a particular dSph is
\begin{equation}
  \Nbar_\DM = \PhiPP \times J(\Delta \Omega) \times (T_\obs\Abareff) \ ,
\end{equation}
where $J (\Delta \Omega)$ is the astrophysical $J$-factor, $T_\obs$ is the observation time, $\Abareff$ is the average effective area of the detector, and $\PhiPP$, a quantity determined only by the DM particle physics model, is given by
\begin{equation}
  \PhiPP = \frac{(\sigma v )_0}{8\pi m_X^2}
  \int_{E_\textrm{th}}^{E_\textrm{max}} dE_\gamma \frac{dN_\gamma}{dE_\gamma}
  \frac{\Aeff(E_\gamma)}{\Abareff} \ ,
\end{equation}
where $m_X$ is the DM mass and $dN_\gamma / dE_\gamma$ is the photon energy spectrum per annihilation.
The effective area of the Fermi-LAT is energy dependent; however, we work in an energy range in which the effective area is approximately constant [$\Aeff(E_\gamma) = \Abareff$] at the few percent level.

The annihilation cross section times relative velocity is often assumed to be constant: $(\sigma v)=(\sigma v)_0$.
To account for a nontrivial dependence on $v$ on the calculation of the $J$-factor (which is determined by the DM velocity distribution~\cite{Boddy:2017vpe}),
we write the annihilation cross section as
\begin{equation}
  \sigma v = (\sigma v)_0 \times S(v) \ ,
\end{equation}
where $S(v)$ is some function of the relative velocity.
The $J$-factor is then
\begin{equation}
  J(\Delta \Omega) = \int_{\Delta \Omega} d\Omega \int d\ell \int d^3 v_1
  f(r(\ell, \Omega), \vec{v}_1) \int d^3 v_2
  f(r(\ell, \Omega), \vec{v}_2) \times
  S(|\vec{v}_1 - \vec{v}_2|) \ ,
\end{equation}
where $\ell$ is the distance along the line-of-sight and $f(r,\vec{v})$ is the DM velocity distribution.
In the limit of $s$-wave annihilation [$S(v)=1$], we recover the standard result for $\sigma v = (\sigma v)_0$: $J = \int_{\Delta \Omega} d\Omega \int d\ell \, \rho^2$.
Note that although dSphs are ideal systems for searches of DM annihilation, this formalism is also applicable for DM decay by substituting $(\sigma v )_0/2m_X \rightarrow \Gamma$ and $J \rightarrow J_D \equiv \int_{\Delta \Omega} d\Omega \int d\ell \, \rho$, where $\Gamma$ is the decay width.

The factors that go into $\Nbar_\DM$ can thus be categorized in the following way:
\begin{enumerate}
\item{$T_\obs \Abareff$ depends on the specifications of the detector.}
\item{$\PhiPP$ depends only the particle physics model for DM annihilation.}
\item{$J$ contains information about the DM distribution, as well as information about the velocity-dependence of the particle physics model for DM annihilation.}
\end{enumerate}
In particular, $\PhiPP$ is completely independent of the choice of target dSph, while $T_\obs \Abareff$ depends on the region of the sky being observed (\textit{i.e.}, the location of the dSph).
On the other hand, $J$ relies on the detailed properties of the target dSph and is subject to significant systematic uncertainty.

The expected total number of photons arising from DM annihilation in a set of dSphs is
\begin{equation}
  \Nbar_\DM^\tot = \PhiPP \times \left[\sum_{i \in \{\textrm{dSph}\}}
    J^i (\Delta \Omega) \times (T_\obs \Abareff)^i \right] \ .
\end{equation}
Our aim is to place a bound on this quantity using Fermi-LAT data.
The data provide $T_\obs \Abareff$ for each dSph, and we use values of $J$ from a variety of previous works.
The bound on $\Nbar_\DM^\tot$ then translates into a bound on $\PhiPP$.
In order to place constraints on $\Nbar_\DM^\tot$, we first need to find the background distributions for the dSphs.


\subsection{Estimating the astrophysical background}
\label{sec:analysis-bkg}

One of the major advantages of using dSphs to search for DM annihilation is their low baryonic content and clean environment.
Well above the Galactic disk, the expected astrophysical contribution to the observed gamma-ray spectrum is from diffuse emission and point sources.
We can choose a region of interest (ROI) around a particular dSph and quantify how likely it is that the number of counts coming from the location of the dSph is or is not consistent with a DM source, given the number of counts in the ROI slightly away from the dSph.
Following Ref.~\cite{GeringerSameth:2011iw}, we find the empirical background distribution for each dSph with the following procedure:
\begin{enumerate}
\item{Choose an ROI, labeled by $i$, that is centered on the dSph with a radius of $10^\circ$ on the sky.}
\item{The number of observed photons, $N_\obs^i$, from the dSph are all those within a radius of $0.5^\circ$ ($\Delta \Omega = 2.4 \times 10^{-4}$) of the dSph's central location.}
\item{Randomly choose $10^5$ sample regions within the ROI of the same size as the target dSph ($0.5^\circ$).}
\item{Reject any sample region whose boundary intersects the border of the ROI or the boundary of a known source region, defined to be within $0.8^\circ$ of a known point source.}
\item{Histogram the number of counts for the surviving sample regions.}
\end{enumerate}
The resulting histogram is the probability mass function $P_\bgd^i (N_\bgd^i)$ for the ROI to contain $N_\bgd^i$ counts in an arbitrary region of $0.5^\circ$.

Increasing the number of sample regions or increasing the size of the source masks has negligible effects on our overall results.
We chose the size of the target and sample regions to be $0.5^\circ$ because many $J$-factor calculations are performed over a cone of radius $0.5^\circ$.
We note that there are certain dSphs for which a known point source is within 1.3$^\circ$, which violates the above criteria for distinguishing the target, background, and known point source regions.
Previous studies~\cite{GeringerSameth:2011iw,GeringerSameth:2014qqa} have included these ``contaminated'' dSphs in their counting analyses, possibly weakening their results.
While we acknowledge this issue, the gain from including more dSphs outweighs the disadvantage of incorporating additional photons whose origin is likely a nearby point source.
Using the contaminated dSphs is acceptable for placing upper limits on DM, but we note that they cannot be used to make a claim for a DM signal.


\subsection{Constraining dark matter}
\label{sec:analysis-DM}

Once we have determined the background distributions for individual dSphs, we convolve these distributions to find the total probability mass function for a set of stacked dSphs:
\begin{equation}
  P_\bgd^\tot (N_\bgd^\tot) \equiv
  \sum_{\sum_i N_\bgd^i = N_\bgd^\tot} \prod_i P_\bgd^i (N_\bgd^i) \ .
\end{equation}
The total number of observed photons is $N_\obs^\tot = \sum_i N_\obs^i$.
For a given expected number of photons arising from DM annihilation, we assume that the actual number of such photons is drawn from a Poisson distribution,
\begin{equation}
  P_\DM^\tot (N_\DM^\tot; \Nbar_\DM^\tot) = e^{-\Nbar_\DM^\tot}
  \frac{(\Nbar_\DM^\tot)^{N_\DM^\tot}}{N_\DM^\tot!} \ .
\end{equation}

The expected total distribution is the convolution of the DM signal and the background.
For an input value of $\Nbar_\DM^\tot$, the probability of producing more than the total observed number of photons $N_\obs^\tot$ from the dSphs is
\begin{equation}
  \sum_{N_\bgd^\tot + N_\DM^\tot > N_\obs^\tot} P_\bgd^\tot (N_\bgd^\tot) \times
  P_\DM^\tot (N_\DM^\tot ; \Nbar_\DM^\tot ) \ .
\end{equation}
Then, the upper bound on the expected number of photons arising from DM annihilation (at confidence level $\beta$), $N_\bound (\beta)$, is given by
\begin{equation}
  \sum_{N_\bgd^\tot + N_\DM^\tot > N_\obs^\tot} P_\bgd^\tot (N_\bgd^\tot) \times
  P_\DM^\tot (N_\DM^\tot ; N_\bound (\beta)) = \beta \ .
\end{equation}
Any model for which $\Nbar_\DM^\tot > N_\bound (\beta)$ may be rejected at the $\beta$ confidence level.
Note that this upper bound on the expected number of total photons arising from DM annihilation is derived entirely from Fermi-LAT data, with no dependence on either the particle physics model or any astrophysical assumptions about the DM velocity or density distribution.

The corresponding upper bound on $\PhiPP$ at $\beta$ confidence level is
\begin{equation}
  \PhiPP^\bound (\beta) \equiv N_\bound (\beta)
  \left[\sum_i J^i \times (T_\obs \Abareff)^i \right]^{-1} \ .
  \label{eqn:PhiPPbound}
\end{equation}
We treat the systematic uncertainties in the $J$-factors following the approach of Ref.~\cite{GeringerSameth:2011iw}.
In particular, $\PhiPP^\bound (\beta)$ is the $\beta$ confidence-level bound on $\PhiPP$ for fixed values of the $J$-factors; it is determined only by the statistical fluctuations in the number of photons produced by DM annihilation in dSphs with those fixed $J$-factors.
A different choice of $J$-factors would yield a different value of $\PhiPP^\bound(\beta)$, and we estimate the astrophysical systematic uncertainty in $\PhiPP^\bound (\beta)$ by determining the range of $\PhiPP^\bound (\beta)$ from varying the values of the $J$-factors within their systematic uncertainties.


\section{Results}
\label{sec:results}

We apply this formalism to the Fermi-LAT Pass 8 data set in the mission elapsed time range of 239557417 to 533867602 seconds.
We incorporate photons in the energy range $1$--$100~\GeV$, with \texttt{evclass=128} and \texttt{evtype=3}.
We set \texttt{\protect{\detokenize{zmax=100}}} and use the filter `\texttt{\protect{\detokenize{(DATA_QUAL>0)&&(LAT_CONFIG==1)}}}'.
To process the Fermi-LAT data, we use the Fermi Science Tools, \texttt{v10r0p5}.

In the following subsections, we verify our methodology, determine $N_\bound(\beta)$ for several different sets of dSphs, and present our constraints on $\Phi_{\textrm{PP}}$.
Finally, we apply this analysis to constrain model parameters in several particle physics scenarios.


\subsection{Comparison to prior results}

In order to verify that our stacking procedure gives reasonable bounds on $\PhiPP$ relative to more complicated analyses, we first reproduce the analysis of Ref.~\cite{GeringerSameth:2011iw}.
They weight events by the signal-to-noise ratio expected from each individual dSph.  In their analysis, an excess event from a dSph with a larger $J$-factor has a greater probability of being a signal event and thus has more constraining power.
We mimic their Pass 7 analysis as closely as possible, with the exception of using a more recent version of Fermi Science Tools, and find $\PhiPP$ using the same $J$-factors from Ref.~\cite{Ackermann:2011wa}.
We find $\PhiPP = 5.54^{+12.11}_{-3.86} \times 10^{-30} ~\cm^3\s^{-1}\GeV^{-2}$ at 95\% C.L. compared to their value of $\PhiPP = 5.0^{+4.3}_{-4.5} \times 10^{-30} ~\cm^3\s^{-1}\GeV^{-2}$.

In the following analysis we opt for the simplest weighting scheme for stacking dSphs, \textit{i.e.}, all events are equally weighted, as described at the beginning of this section.
This stacking procedure yields $\PhiPP = 6.62^{+9.38}_{-4.27} \times 10^{-30} ~\cm^3\s^{-1}\GeV^{-2}$ at 95\% C.L.
Although the stacking bound is weaker, it is consistent with the signal-to-noise bound, given the uncertainties in the $J$-factors.


\subsection{Determination of $N_\bound (\beta)$ from Fermi-LAT data}

We consider five sets of dSphs as detailed in Table~\ref{tab:targets} in the Appendix:
\begin{enumerate}
\item{\textit{Set 1:} The set of 45 objects considered in Ref.~\cite{Fermi-LAT:2016uux}, which includes 28 confirmed dSphs, 13 likely galaxies, and 4 ambiguous systems.}
  \begin{enumerate}
  \item{\textit{Set 1a:} The subset that includes only the 28 confirmed dSphs.}
  \item{\textit{Set 1b:} The subset that includes the 28 confirmed dSphs and the 13 likely galaxies.}
  \item{\textit{Set 1c:} The subset that includes 27 dSphs for which the $0.5^\circ$ radius around the central location of each dSph does not intersect the $0.8^\circ$ mask around nearby point sources.}
  \end{enumerate}
\item{\textit{Set 2:} The set of 27 dSphs for which $s$-wave $J$-factors have been calculated in Ref.~\cite{Evans:2016xwx}.}
\item{\textit{Set 3:} The set of 24 dSphs for which $J$-factors for non-spherical halos have been calculated in Ref.~\cite{Hayashi:2016kcy}.}
\item{\textit{Set 4:} The set of 7 dSphs for which $J$-factors modified for foreground effects have been calculated in Refs.~\cite{Ichikawa:2016nbi,Ichikawa:2017rph}.}
\item{\textit{Set 5:} The set of 5 dSphs considered in Ref.~\cite{Boddy:2017vpe}, for which Sommerfeld-enhanced $J$-factors have been calculated in the Coulomb limit.}
\end{enumerate}
Each of these is a different set of objects, although many dwarfs appear in multiple sets.
The differences between these sets lie in one's assessment of which objects are actually dwarf spheroidal galaxies (and thus should be used in a search for DM annihilation),
in the possibility of background contamination from point sources,
and in assumptions about how one computes the $J$-factor (including how to treat systematic uncertainties, assumptions about the DM mass distribution, and the effect of the velocity-distribution on DM annihilation).
These assumptions thus determine which of the above sets of objects are appropriate for a DM search.
Given that choice of the set of objects, the quantity that is relevant to a search for DM annihilation is $N_\bound(\beta)$, the upper bound (at confidence level $\beta$) on the total expected number of photons arising from DM annihilation in that set of objects.
This quantity encapsulates everything that one needs to know from Fermi-LAT photon data.

\begin{figure}[t]
  \centering
  \includegraphics[scale=0.6]{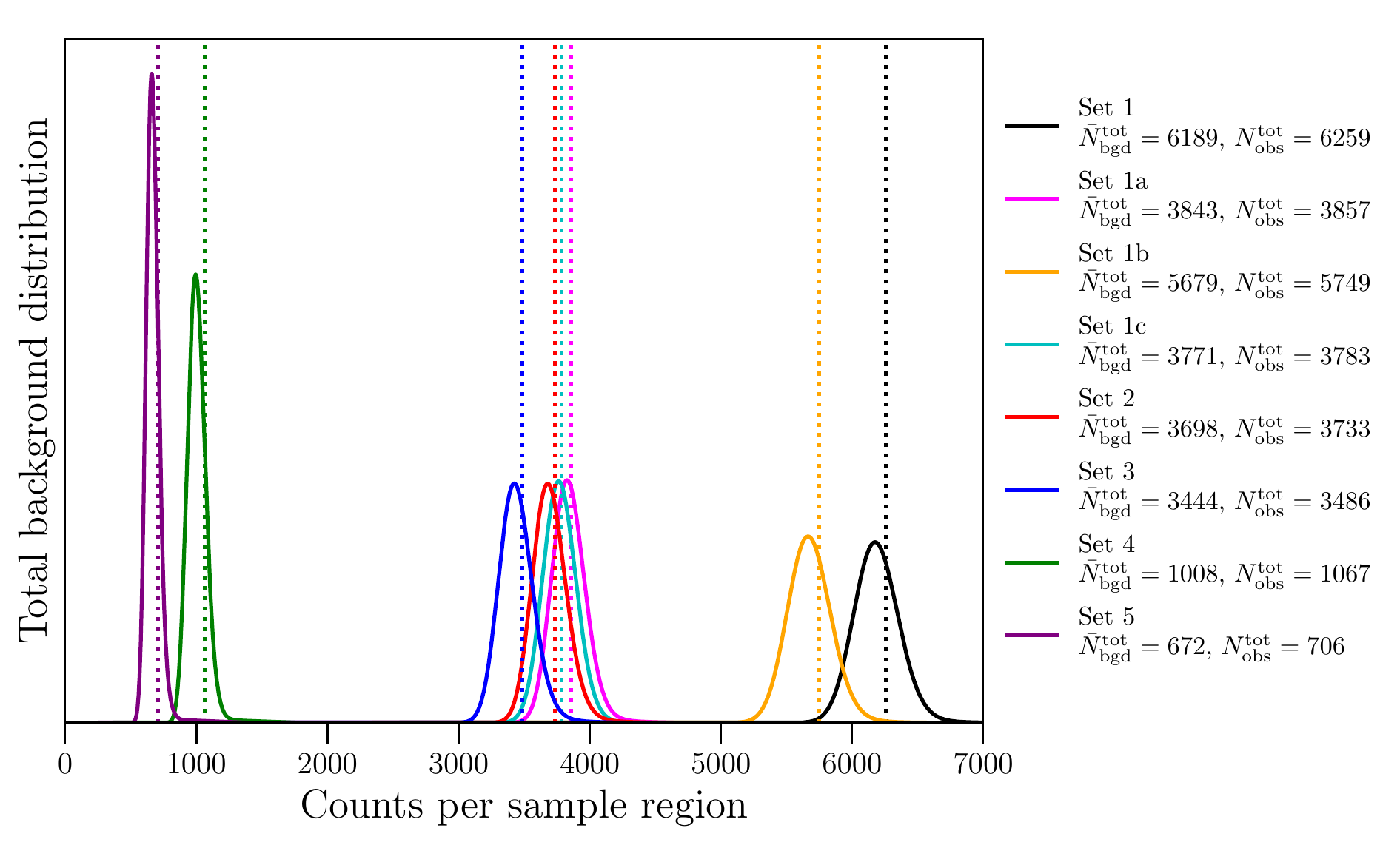}
  \caption{Total background distribution, $P_\bgd^\tot (N_\bgd^\tot)$, for the sets of astrophysical objects discussed in the text (solid lines).
  The actual numbers of counts observed from each set of objects are marked by dotted lines.
  The curves from left to right are for \textit{Set 5}, \textit{Set 4}, \textit{Set 3}, \textit{Set 2}, \textit{Set 1c}, \textit{Set 1a}, \textit{Set 1b}, and \textit{Set 1}.}
  \label{fig:bkg-total}
\end{figure}

In Fig.~\ref{fig:bkg-total}, we plot the total background distribution for each set of dSphs, as well as the number of photons observed.
For each set, the expected total number of background counts ($\overline N_\bgd^\tot$) and the actual number of photons observed ($N_\obs^\tot$) are given in the figure legend.
The background photon distributions for each of the individual dSphs is provided in Fig.~\ref{fig:pmf} in the Appendix and in the supplementary file.
In Fig.~\ref{fig:Nbound}, we plot $N_\bound (\beta)$ for each set of dSphs.
Note that the normalizations of the background distributions do depend implicitly on the Fermi-LAT exposures on each dwarf, which are listed in Table~\ref{tab:targets} in the Appendix.
Although the choice of the appropriate set of dSphs may be motivated by assumptions about astrophysics, $N_\bound(\beta)$ itself is entirely independent of any assumptions about DM physics.
For example, to constrain a model of Sommerfeld-enhanced DM annihilation, the dSphs given in \textit{Set 5} should be used, because Sommerfeld-enhanced $J$-factors have been computed for those objects.
If Sommerfeld-enhanced $J$-factors are eventually determined for all of the objects in the larger \textit{Set 1}, then one may instead use that set of objects; the only input needed from Fermi-LAT photon data would be $N_\bound^{Set~1} (\beta)$ already presented above.

Indeed, Sommerfeld-enhanced $J$-factors have recently been computed in Ref.~\cite{Bergstrom:2017ptx} for a set of 20 dSphs, though using a different methodology than that used in Ref.~\cite{Boddy:2017vpe}.
Although this set of 20 dSphs is not one of those for which we have plotted $N_\bound (\beta)$, it is possible to compute $N_\bound (\beta)$ for \textit{any} set of the 47 objects of Fig.~\ref{fig:pmf}, using the background distributions and the numbers of observed counts found therein, as well as the formulae in Section~\ref{sec:analysis-DM}.

\begin{figure}[t]
  \centering
  \includegraphics[scale=0.6]{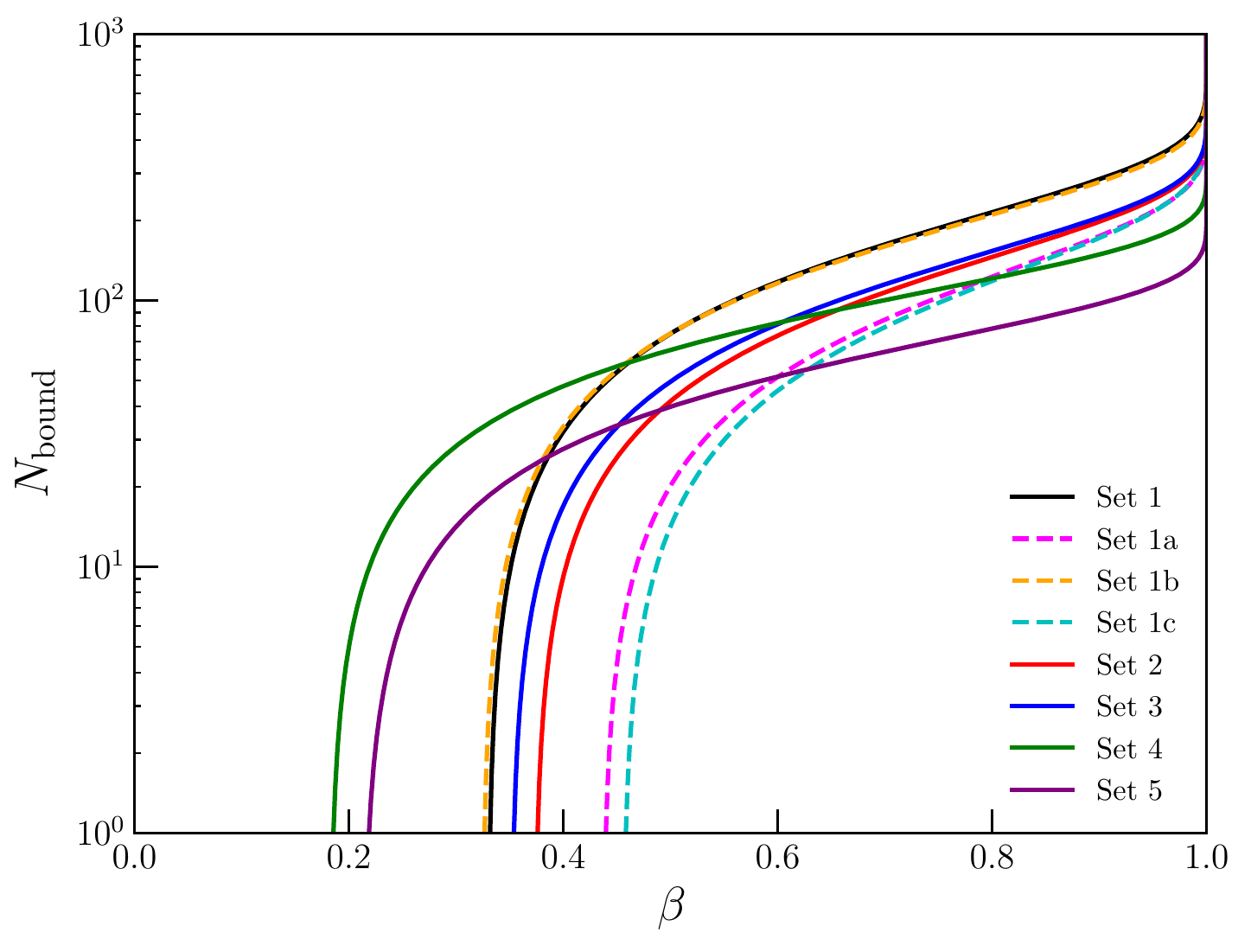}
  \caption{$N_\bound(\beta)$ for each of the sets of astrophysical objects discussed in the text.}
  \label{fig:Nbound}
\end{figure}


\subsection{Constraints on $\PhiPP$}

\begin{figure}[t]
  \centering
  \includegraphics[width=0.45\textwidth]{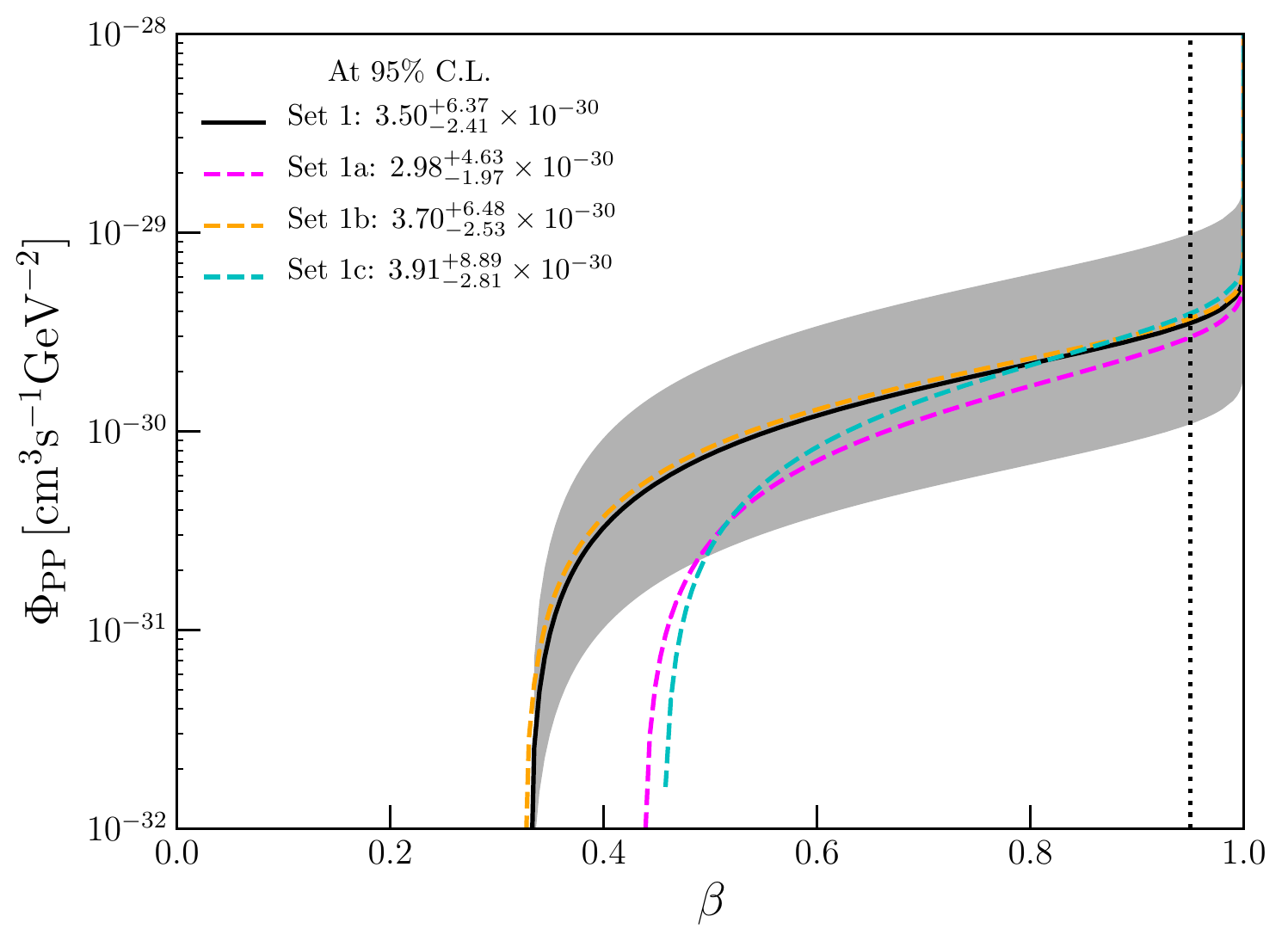}
  \includegraphics[width=0.45\textwidth]{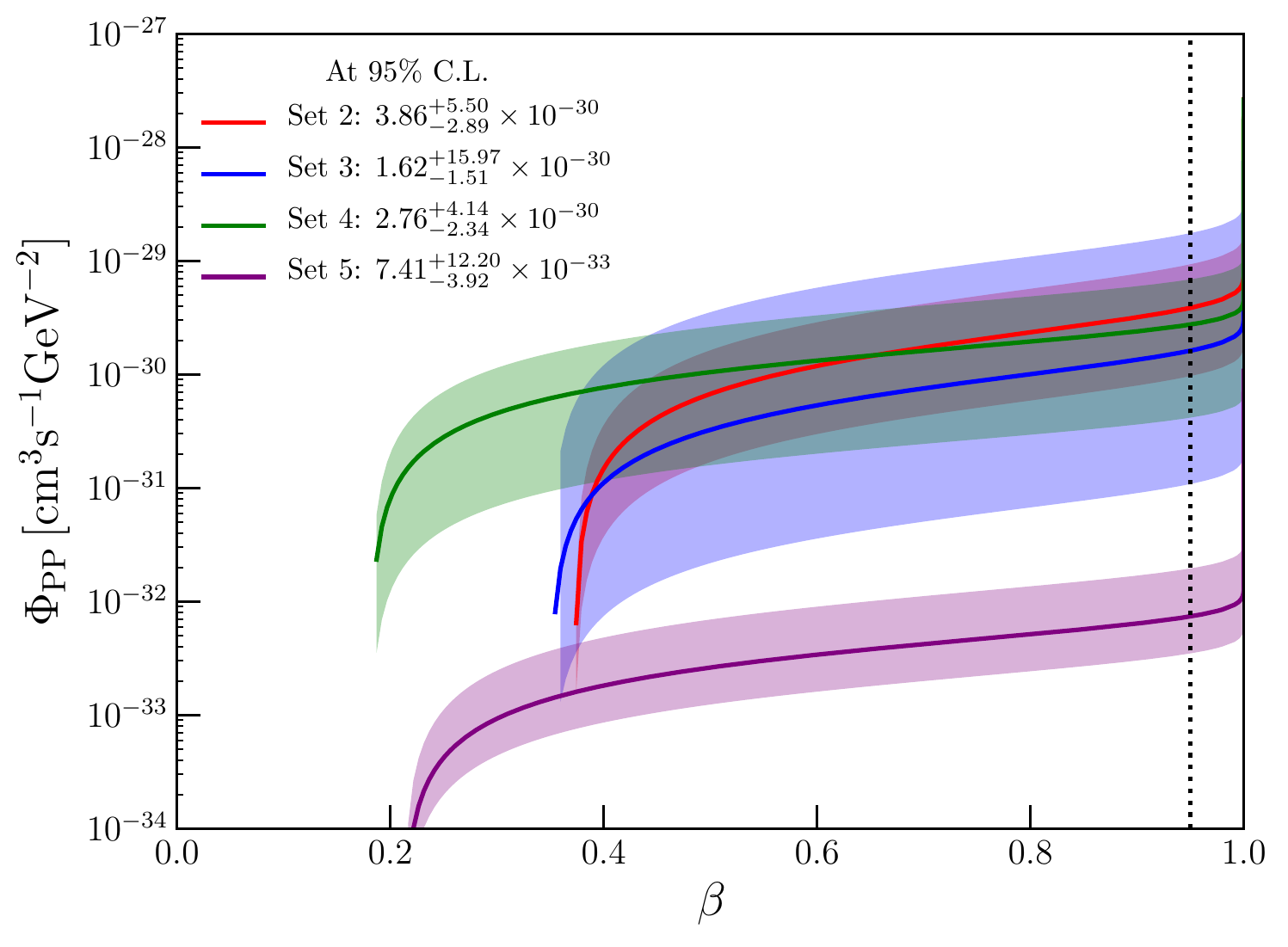}
  \caption{$\PhiPP^\bound (\beta)$ for the data sets discussed in the text.
    For each set of target objects, the solid line corresponds to central value of the $J$-factor for each object, while the edges of the band correspond to the $\pm 1\sigma$ variation of the $J$-factors for all objects.
    In the left panel the $\pm 1\sigma$ variation is only shown for \textit{Set 1}.
    The vertical dotted lines mark the 95\% C.L.}
  \label{fig:PhiPPbound}
\end{figure}

To transform $N_\bound(\beta)$ into a constraint on $\PhiPP$, we must now plug in specific $J$-factors.
We consider the data sets of the previous subsection with their associated $J$-factors.
In Fig.~\ref{fig:PhiPPbound}, we plot $\PhiPP^\bound (\beta)$ as a function of $\beta$ for each of these sets.
In each case, the width of the band arises from varying all $J$-factors through their $1\sigma$ uncertainties.
Using Eq.~\eqref{eqn:PhiPPbound} and the exposures given in Table~\ref{tab:targets} in the Appendix, $\PhiPP^\bound (\beta)$ can be rescaled appropriately for any determination of the relevant $J$-factors.

Note that for \textit{Set 5}, the Sommerfeld-enhanced $J$-factors were computed assuming that the dark fine structure constant is $\alpha_X = 0.01$, and in the limit of a Coulomb-like interaction, the Sommerfeld-enhanced $J$-factor is proportional to $\alpha_X$~\cite{Boddy:2017vpe}.
For a different choice of $\alpha_X$, $\Phi_{\rm PP}^{\bound, Set~5} (\beta)$ should be rescaled by a factor $0.01/\alpha_X$.

\subsection{Constraints on particle physics parameters}

Finally, we translate constraints on $\PhiPP$ into constraints on DM parameters, for several choices of interaction models.
For the purpose of illustration in this subsection, we focus on obtaining constraints on different particle physics models, while making nominal assumptions about DM astrophysics.

We consider the following particle physics scenarios:
\begin{enumerate}
\item{\textit{Particle Physics Scenario 1}: DM with mass $m_X$ annihilates with a total $s$-wave cross section $(\sigma v)_0$ to a two-body final state.
  We consider $\bar \tau \tau$, $\bar b b$, $W^+ W^-$and $\bar \mu \mu$ final states, each with 100\% branching fraction, and the dSphs and associated $J$-factors of \textit{Set 1}.}
\item{\textit{Particle Physics Scenario 2}: DM with mass $m_X$ self-interacts through a long-range Yukawa force with coupling strength $\alpha_X = 10^{-2}$ and annihilates through a contact interaction with cross section $(\sigma v)_0$.
  DM annihilation is thus Sommerfeld-enhanced.
  We consider $\bar \tau \tau$, $\bar b b$, $W^+ W^-$and $\bar \mu \mu$ final states, each with 100\% branching fraction, and the dSphs and associated $J$-factors of \textit{Set 5}.}
\item{\textit{Particle Physics Scenario 3}: DM with mass $m_X$ annihilates with a total $s$-wave cross section $(\sigma v)_0$ to a three-body final state $\bar \mu \mu \gamma$ via internal bremsstrahlung.
  This situation occurs if DM annihilation to $\bar \mu \mu$ is $p$-wave suppressed and internal bremsstrahlung is the dominant annihilation channel.
  We consider the model presented in Ref.~\cite{Kumar:2016cum}, with two charged mediators of masses $m_1$ and $m_2$ and left-right mixing angle $\theta_{LR}$, and the dSphs and associated $J$-factors of \textit{Set 1}.}
\item{\textit{Particle Physics Scenario 4}: DM with mass $m_X$ annihilates with total $s$-wave cross section $(\sigma v)_0$ to a pair of intermediate particles $\phi$ (of mass $m_\phi$), each of which decays to two photons ($XX \rightarrow \phi \phi \rightarrow 4\gamma$).
  We consider the dSphs and associated $J$-factors of \textit{Set 1}.}
\item{\textit{Particle Physics Scenario 5}: DM consists of a Dynamical Dark Matter (DDM) ensemble~\cite{DDM2}, the lightest component of which has mass $m_0$ and annihilates with cross section $(\sigma v)_0$ to a pair of intermediate particles $\phi$, each of which decays to two photons ($X_i X_i \rightarrow \phi \phi \rightarrow 4\gamma$)~\cite{Boddy:2016hbp}.
  The heavier components of the ensemble (with mass $m_n$) annihilate to the same final state, but with a rate which scales as $\propto (m_n / m_0)^\xi$, where $\xi$ is a parameter.
  We consider the dSphs and associated $J$-factors of \textit{Set 1}.}
\end{enumerate}

\begin{figure}[t]
  \centering
  \includegraphics[scale=0.9]{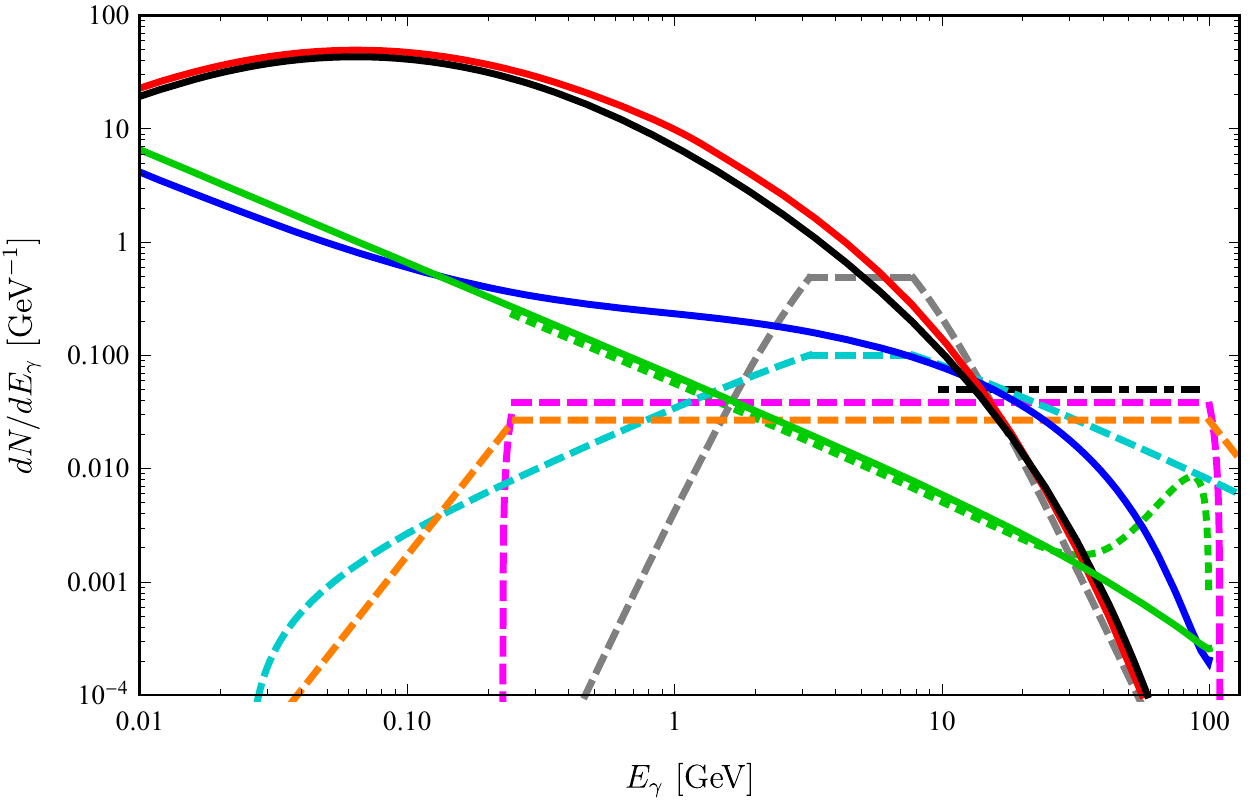}
  \caption{Photon energy spectra for the final states $\bar \tau \tau$ (blue solid), $\bar b b$ (red solid), $\bar \mu \mu$ (green solid), $W^+ W^-$ (black solid), $\bar \mu \mu \gamma$ via internal bremsstrahlung (green dotted), $XX \rightarrow \phi \phi \rightarrow 4\gamma$ (black dotted), and for four DDM scenarios described in the text (pink, grey, cyan, and orange dashed lines).
    For all scenarios other than DDM, $m_X=100~\GeV$.}
  \label{fig:dNdxPlot}
\end{figure}

We plot representative photon spectra for these scenarios in Fig.~\ref{fig:dNdxPlot} assuming that the primary contributions to the photon flux are final state radiation and secondary decays, and that propagation effects in dSphs are negligible.
Each spectrum is normalized to the average number of photons produced per annihilation, and in all cases of single-component DM, we take \mbox{$m_X = 100~\GeV$}.
\textit{Particle Physics Scenario 1} is widely studied, and for this case we plot spectra obtained from the tools provided in Ref.~\cite{Cirelli:2010xx}, for the final states  $\bar \tau \tau$ (blue solid), $\bar b b$ (red solid), $\bar \mu \mu$ (green solid), and $W^+ W^-$ (black solid).
\textit{Particle Physics Scenarios 2} yields the same photon spectra as \textit{Particle Physics Scenario 1}, but with different $J$-factors.
The spectrum for \textit{Particle Physics Scenario 3} is plotted as a green dashed curve  assuming the masses of the charged mediators are $m_1=120~\GeV$ and $m_2=450~\GeV$ and the mediator left-right mixing angle is $\theta_{LR} = 0$.
The dotted black curve shows an example photon spectrum for \textit{Particle Physics Scenario 4}, under the assumption that the mediator mass is $m_\phi = 60~\GeV$.
Finally, the spectra for the DDM \textit{Particle Physics Scenario 5} correspond to $m_\phi =10~\GeV$, and  $(m_0, m_\textrm{max}, \xi)$ = $(11~\GeV, 1000~\GeV, -5)$ (grey dashed), $(11~\GeV, 1000~\GeV, -1)$ (cyan dashed), $(100~\GeV, 110~\GeV, -1)$ (magenta dashed) and  $(100~\GeV, 10000~\GeV, -3)$ (orange dashed).
This scenario yields photon energy spectra that are invariant under the transformation $E_\gamma \rightarrow m_\phi^2 /4E_\gamma$~\cite{Agashe:2012bn}.

\begin{figure}[t]
  \centering
  \includegraphics[scale=0.9]{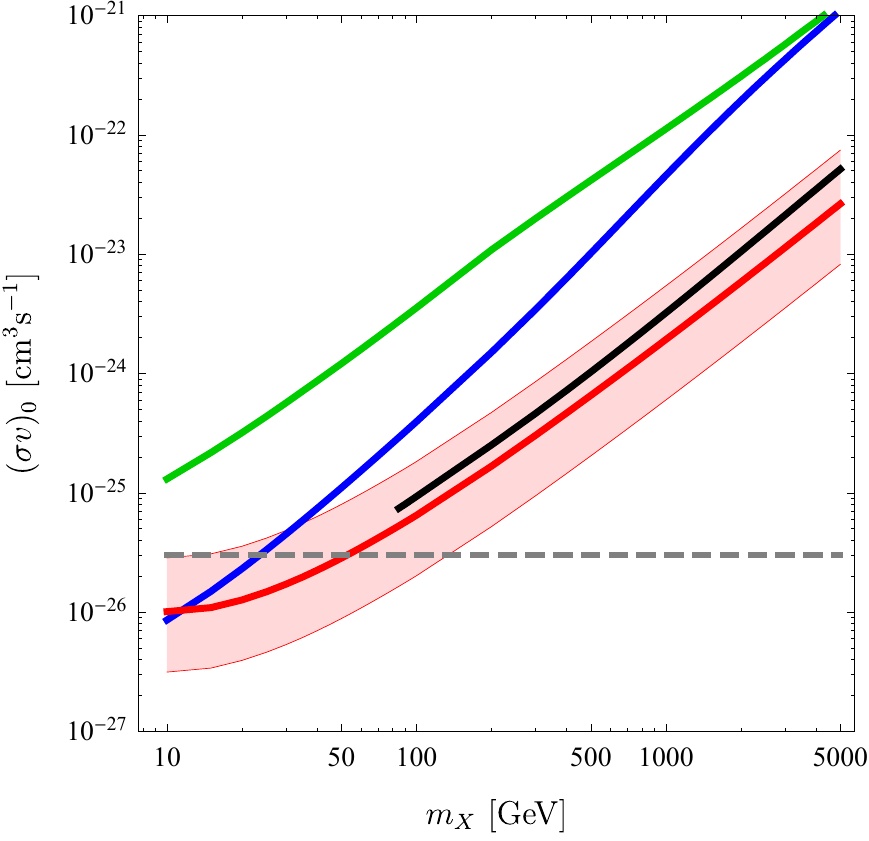} \hspace{5mm}
    \includegraphics[scale=0.9]{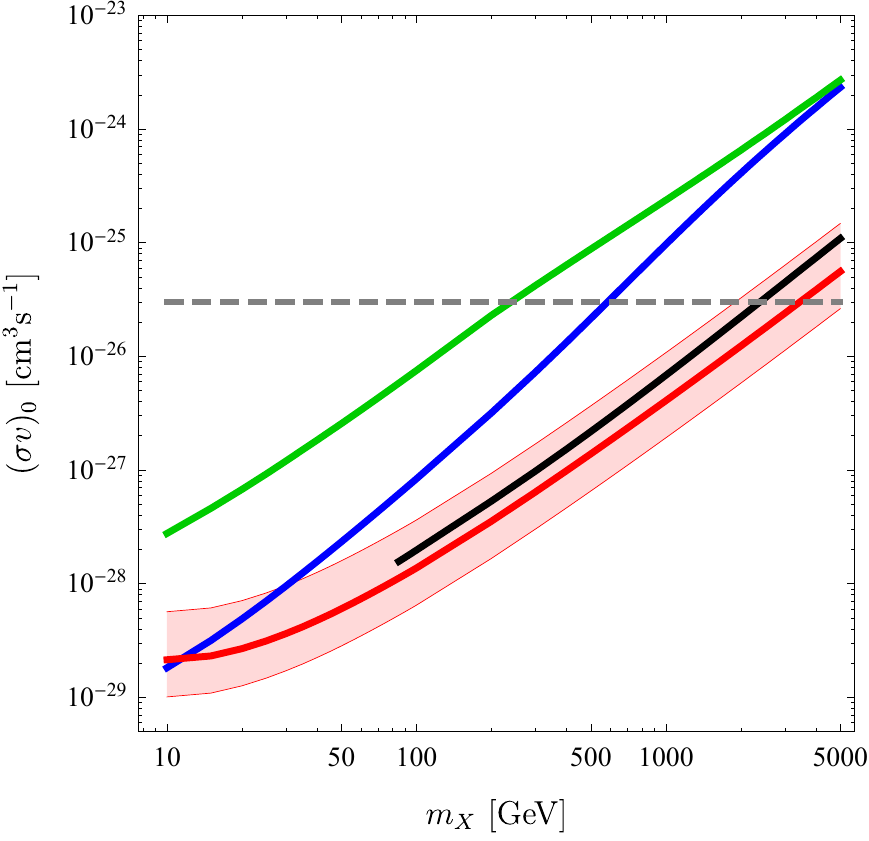}
  \caption{The 95\% C.L. bounds on $(\sigma v )_0$ as a function of $m_X$ for \textit{Particle Physics Scenario 1} (left) and the Sommerfeld-enhanced \textit{Particle Physics Scenario 2} (right), assuming the final state (with 100\% branching fraction) is $\bar \tau \tau$ (blue), $\bar b b$ (red), $\bar \mu \mu$ (green) and $W^+ W^-$ (black).
    For the case of annihilation to $\bar b b$ in each panel, we show the effect on the variation in the 95\%~C.L. limits due to the 1$\sigma$ variation in $J$-factors for all objects considered, as presented in Fig.~\ref{fig:PhiPPbound}, with the red shading.
    The grey dashed line in each panel indicates a cross section of $(\sigma v )_0=3\times 10^{-26}$ cm$^3$s$^{-1}$.}
  \label{fig:Standard2BodyLimits}
\end{figure}

In Fig.~\ref{fig:Standard2BodyLimits}, we plot 95\% C.L. bounds on $(\sigma v )_0$ as a function of $m_X$ for \textit{Particle Physics Scenario 1} (left) and the Sommerfeld-enhanced \textit{Particle Physics Scenario 2} (right), assuming the final state (with 100\% branching fraction) is $\bar \tau \tau$ (blue), $\bar b b$ (red), $\bar \mu \mu$ (green), and $W^+ W^-$ (black).
For the case of annihilation to $\bar b b$, in each panel we show the effect on the 95\% C.L. limits due to the 1$\sigma$ variation in $J$-factors for all objects considered (as presented in Fig.~\ref{fig:PhiPPbound}), with the red shading.
For reference, the grey dashed line in each panel indicates a cross section of $(\sigma v )_0=3\times 10^{-26}$~cm$^3$s$^{-1}$.
The limits are expectedly stronger for the Sommerfeld-enhanced scenario than in the absence of a Sommerfeld enhancement.
Focusing on \textit{Particle Physics Scenario 1} (left), one can compare the limits presented here to those presented in Ref.~\cite{Ackermann:2015zua}; for $m_X = 10~\GeV$, the limits presented here are weaker by a factor of $\sim 2-5$, which is less than the systematic uncertainty of this analysis.
Of course, a direct comparison of these methodologies is not readily possible, since the set of targets for the two analyses are different.
However, it is unsurprising that a dedicated study of a particular particle physics model proves more constraining than a generic search.
The more interesting application of this method is to models for which current constraints are inapplicable, as shown in the right panel.

\begin{figure}[t]
  \centering
  \includegraphics[scale=0.9]{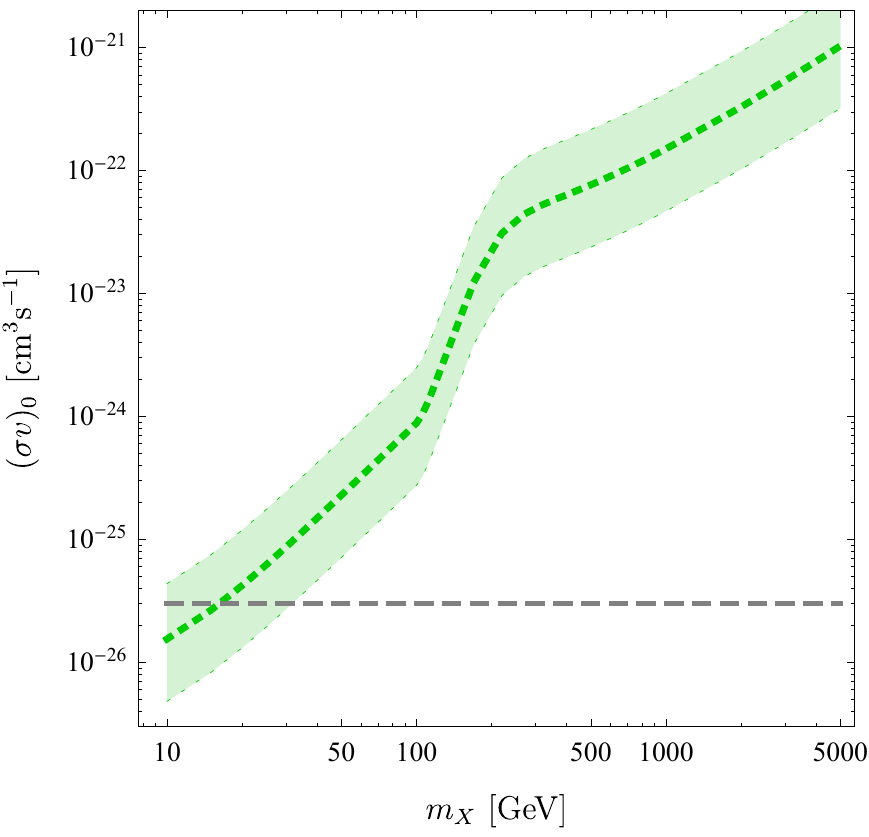}
 \caption{The 95\%~C.L. bounds on $(\sigma v )_0$ as a function of $m_X$, for \textit{Particle Physics Scenario 3}, assuming the final state (with 100\% branching fraction) is $\bar \mu \mu$.
   The mediator masses are $m_1=1.2m_X$ and $m_2=4.5m_X$, and the mixing angle is $\theta_{LR}=0$.
   We also show the effect on the variation in the 95\%~C.L. limits due to the 1$\sigma$ variation in $J$-factors for all objects considered, as presented in Figure~\ref{fig:PhiPPbound}, with the shaded region.
   The grey dashed line indicates a cross section of $(\sigma v )_0=3\times 10^{-26}$ cm$^3$s$^{-1}$.}
 \label{fig:IB}
\end{figure}

For \textit{Particle Physics Scenario 3}, we first consider the case in the absence of left-right mixing (\textit{i.e.},~$\theta_{LR}=0$), which exhibits a substantial bump in the photon spectrum near the DM mass due to virtual internal bremsstrahlung (VIB).
In the limit in which the lightest charged mediator is nearly degenerate with the DM ($m_1 \sim m_X$), the photon is very hard and the spectrum is not very different from that of a line.
On the other hand, if $m_1 \gg m_X$, the effects of VIB are largely irrelevant.
We focus on the intermediate case, $m_1 \gtrsim m_X$, for which the spectral shape is not well approximated by typical spectra utilized in dSph searches.
In Fig.~\ref{fig:IB}, we plot the bounds on $(\sigma v )_0$ for \textit{Particle Physics Scenario 3}, assuming $\theta_{LR} =0$ and for mediator masses $m_1=1.2m_X$ and $m_2=4.5m_X$.

\begin{figure}[t]
  \centering
    \includegraphics[scale=0.9]{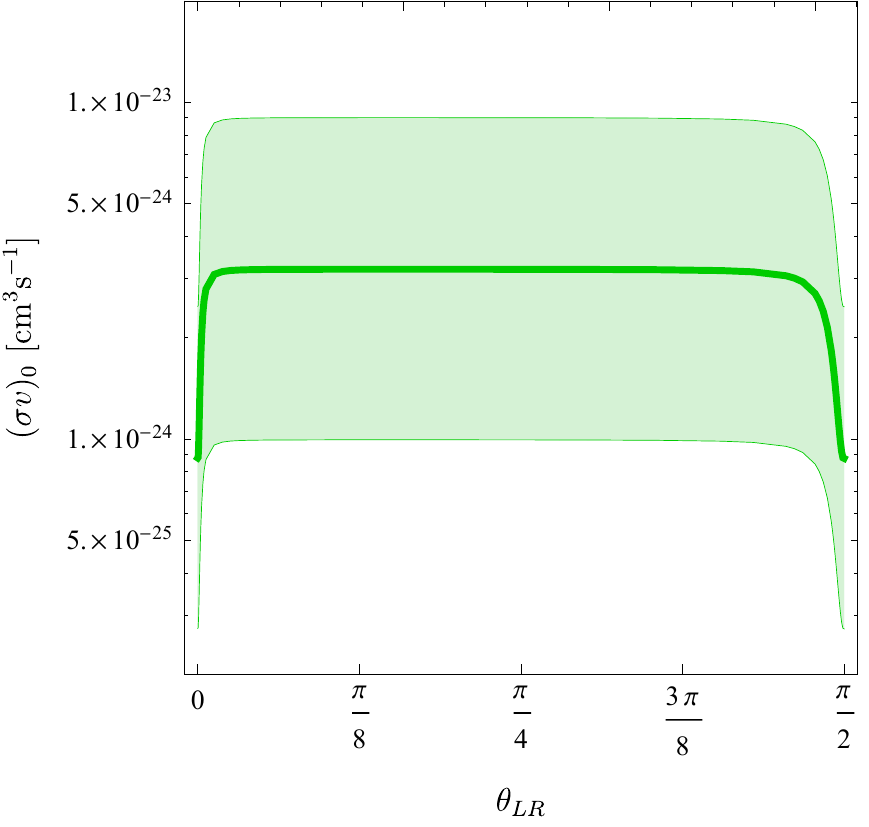} \hspace{5mm}
      \includegraphics[scale=0.9]{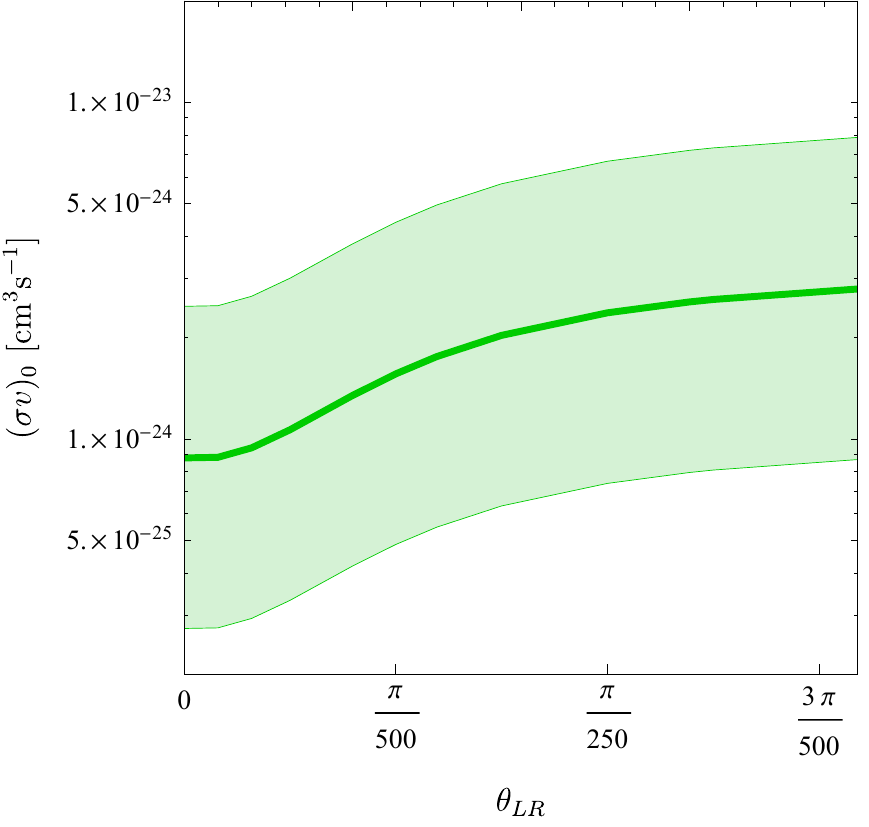}
  \caption{The 95\% C.L. bounds on $(\sigma v )_0$ as a function of mixing angle $\theta_{LR}$ for \textit{Particle Physics Scenario 3}, where we take the DM mass to be \mbox{$m_X=100~\GeV$} and the scalar mediator masses to be $m_1=120~\GeV$ and $m_2 = 450~\GeV$.
    In the left panel, we show the range of $\theta_{LR}$ between 0 and $\pi/2$, while in the right panel we focus on small $\theta_{LR}$.
    We also show the effect on the variation in the 95\%~C.L. limits due to the 1$\sigma$ variation in $J$-factors for all objects considered, as presented in Figure~\ref{fig:PhiPPbound}, with the shaded region in each panel.}
  \label{fig:IBangle}
\end{figure}

In Fig.~\ref{fig:IBangle}, we again consider \textit{Particle Physics Scenario 3} with mediator masses $m_1=1.2m_X$ and $m_2=4.5m_X$, but now with fixed $m_X=100$ GeV.  We plot the bounds on $(\sigma v )_0$ as a function of $\theta_{LR}$~\cite{Kumar:2016cum}.
In the left panel, we show the full range of $\theta_{LR}$ between 0 and $\pi/2$; while in the right panel, we consider small $\theta_{LR}$ where the effect of left-right mixing is substantial.
For $\theta_{LR}$ near 0 or $\pi/2$, the VIB bump is substantial and dependent on the value of $\theta_{LR}$, leading to a $\theta_{LR}$-dependent limit.
For moderate values of $\theta_{LR}$, where the limit is flat in the left panel, the photon spectrum does not exhibit a substantial VIB bump and is therefore independent of $\theta_{LR}$.

\begin{figure}[t]
  \centering
  \includegraphics[scale=0.9]{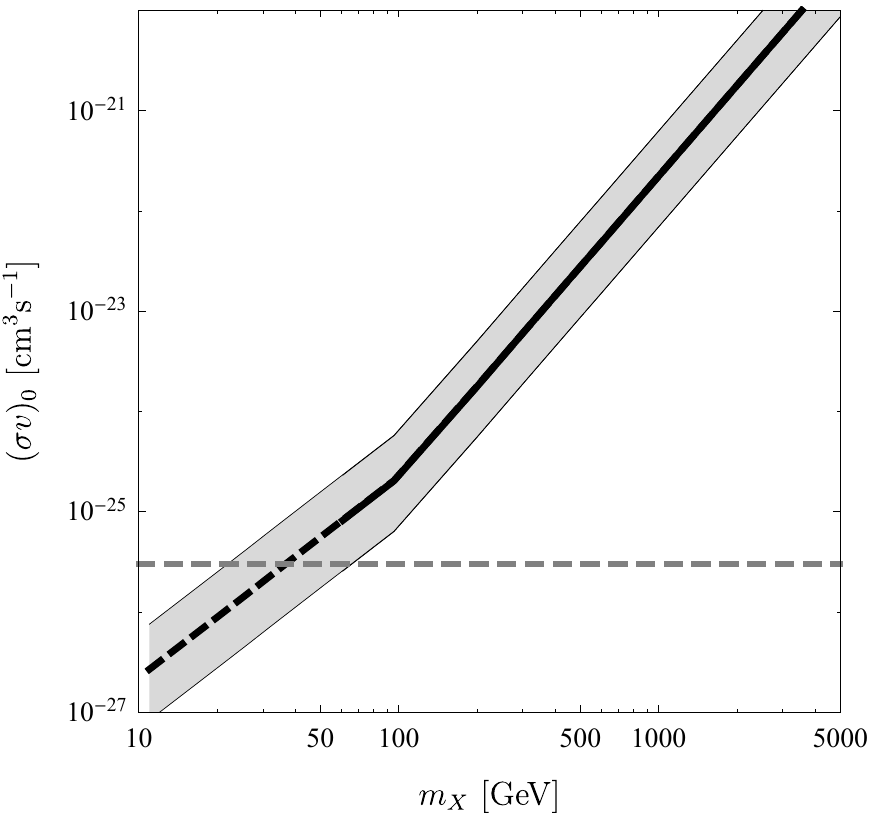}
  \caption{The 95\% C.L. bounds on $(\sigma v )_0$ as a function of $m_X$, for \textit{Particle Physics Scenario 4}, assuming $m_\phi = 10~\GeV$ (dashed) and $60~\GeV$ (solid).
    We also show the effect on the variation in the 95\% C.L. limits due to the 1$\sigma$ variation in $J$-factors for all objects considered, as presented in Fig.~\ref{fig:PhiPPbound}, with the shaded region.
    The grey dashed line indicates a cross section of $(\sigma v )_0=3\times 10^{-26}$ cm$^3$s$^{-1}$.}
  \label{fig:BoxSpectrumLimits}
\end{figure}

In Fig.~\ref{fig:BoxSpectrumLimits}, we plot 95\% C.L. bounds on $(\sigma v )_0$ as a function of $m_X$, for \textit{Particle Physics Scenario 4}, assuming $m_\phi = 10~\GeV$ (dashed) and $60~\GeV$ (solid).

\begin{figure}[t]
  \centering
  \includegraphics[scale=0.97]{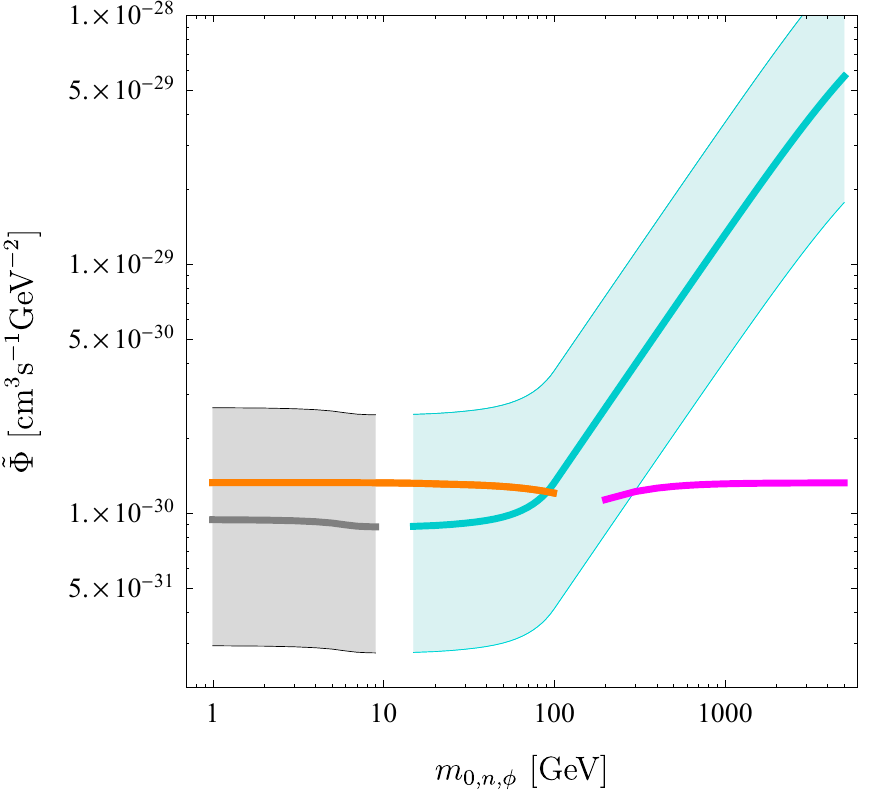}
 \caption{The 95\% C.L. bounds on $\tilde \Phi$ for \textit{Particle Physics Scenario 5}, with $\xi =-3$.
   We plot constraints for four specific models: $(m_\phi, m_0, m_\textrm{max})=(10~\GeV, m, 10000~\GeV)$ (cyan), ($10~\GeV, 100~\GeV, m)$ (magenta), $(m, 10~\GeV, 10000~\GeV)$ (grey), and $(m, 100~\GeV, 10000~\GeV)$ (orange), where $m$ is the quantity plotted on the $x$-axis.
   We also show the effect on the variation in the 95\% C.L. limits due to the 1$\sigma$ variation in $J$-factors for all objects considered, as presented in Fig.~\ref{fig:PhiPPbound}, with the shaded regions surrounding the grey and cyan curves.}
  \label{fig:DDMLimits}
\end{figure}

For \textit{Particle Physics Scenario 5}, we plot
\begin{equation}
  \tilde \Phi \equiv \PhiPP^\bound (\beta =0.95) \times
  \left[\frac{1}{4}\int_{E_\textrm{th}}^{E_\textrm{max}} dE_\gamma
    \frac{dN_\gamma}{dE_\gamma} \frac{\Aeff(E_\gamma)}{\Abareff}  \right]^{-1} \ ,
\end{equation}
 in Fig.~\ref{fig:DDMLimits}.
This quantity, multiplied by the $J$-factor, is the 95\% C.L. bound on the total photon flux at the Fermi-LAT arising from DM annihilation.
If $\Delta m$ is a constant mass splitting between successive DM components, and if $m_0$, $\Omega_0$ and $(\sigma v)_0$ are the mass, abundance, and annihilation cross section of the lightest component, respectively, then we find
\begin{equation}
  \tilde \Phi
  = \frac{(\sigma v)_0}{8\pi m_0^2} \frac{\Omega_0^2}{\Omega_\tot^2}
  \frac{m_0}{(\xi +1) \Delta m}
  \left[\left(\frac{m_\textrm{max}}{m_0} \right)^{\xi+1} -1 \right] \ ,
\end{equation}
where $m_\textrm{max}$ is the mass of the heaviest DM component.
We set $\xi = -3$, and determine a 95\% C.L. bound on $\tilde \Phi$ for a model parameterized by $(m_\phi, m_0, m_\textrm{max})=(10~\GeV, m, 10000~\GeV)$ (cyan), ($10~\GeV, 100~\GeV, m)$ (magenta), $(m, 10~\GeV, 10000~\GeV)$ (grey), and $(m, 100~\GeV, 10000~\GeV)$ (orange), where $m$ is the quantity plotted on the $x$-axis.

\section{Conclusions}
\label{sec:conclusions}

We have described a formalism for deriving model-independent constraints on the number of photons produced by DM annihilation in a set of dwarf spheroidal galaxies.
Our approach differs from previous attempts in that our constraints are independent of both the DM particle physics model and the DM astrophysics.
Essentially, once the number of background photons is estimated by using data taken slightly off-axis, the number of photons originating from DM annihilation can be statistically constrained, independent of any assumptions about how the DM actually produces those photons.
Although such a general search is indeed less powerful than a targeted search strategy for any particular model, the loss in constraining power is not dramatic.

With increasingly diverse models of DM being considered, the utility of a model-independent constraint on DM annihilation in dSphs is clear.
Since models with multibody annihilation final states, with final-state cascades, with multi-component DM, etc., have gained popularity, dSph analyses targeted towards particular sets of photon spectra are not generally applicable to a specific model of interest.
Similarly, not only is there significant uncertainty in the standard $J$-factors applicable for $s$-wave annihilation, but also uncertainty as to whether this is even a correct type of $J$-factor to apply.
If DM decays, or if DM annihilation has nontrivial velocity-dependence, then the modified $J$-factors can be very different from the standard $J$-factors.
In such cases, an analysis which weights the statistical power of photons based on a putative set of $J$-factors would again be inapplicable.

\vskip 0.1in
{\bf Acknowledgements.}
We are grateful to L.~Strigari for useful discussions.
This work is supported in part by NSF~CAREER Grant No.~PHY-1250573, DOE Grant No.~DE-SC0010504, and NSF Grant No.~PHY-1720282.
K.B., J.K. and P.S. thank CETUP* for its hospitality while this work was in progress.
D.M.~thanks the Aspen Center for Physics (which is supported by NSF Grant No.~PHY-1607611) for its hospitality while this work was in progress.

\newpage
\appendix*

\section{Astrophysical and Detector Parameters}

\begin{figure}[h]
  \centering
  \includegraphics[scale=0.7]{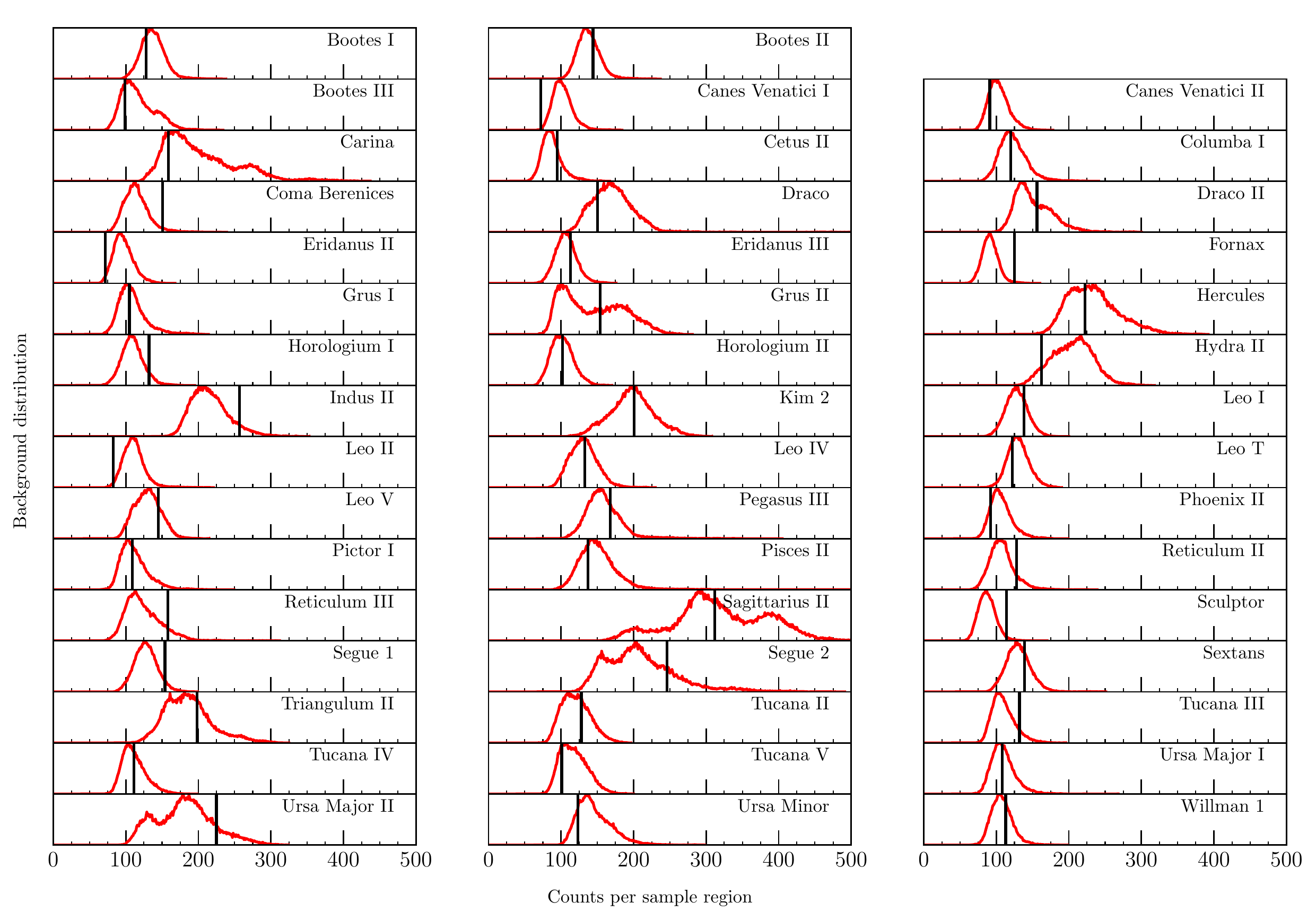}
  \caption{Background distribution (red) for each dwarf, and the observed number of counts (dashed black) from the central region of the ROI.}
  \label{fig:pmf}
\end{figure}

\begin{table}
  \begin{tabular}{l|c|c|c|cccccccc}
    \hline
    \hline
    Name & $\Abareff T_\obs$ & $\overline N_\bgd$ & $N_\obs$ & \multicolumn{8}{c}{$\log_{10}(J/[\GeV^2/\cm^5])$} \\
    & $[\cm^2\s]$ & & & Set 1 & a & b & c & Set 2 & Set 3 & Set 4 & Set 5 \\
    \hline
    Bootes I          & 4.042e+11 &  137 &  128 & $18.2^{+0.4}_{-0.4}$    & \checkmark              & \checkmark              & \checkmark              & $16.65^{+0.64}_{-0.38}$ & $16.95^{+0.53}_{-0.40}$ & -                       & -                       \\
    Bootes II         & 4.012e+11 &  138 &  144 & $18.9^{+0.6}_{-0.6}$    & \checkmark              & \checkmark              & -                       & -                       & -                       & -                       & -                       \\
    Bootes III        & 4.197e+11 &  117 &   99 & $18.8^{+0.6}_{-0.6}$    & \checkmark              & \checkmark              & \checkmark              & -                       & -                       & -                       & -                       \\
    Canes Venatici I  & 4.270e+11 &  102 &   72 & $17.4^{+0.3}_{-0.3}$    & \checkmark              & \checkmark              & \checkmark              & $17.27^{+0.11}_{-0.11}$ & $16.92^{+0.43}_{-0.26}$ & -                       & -                       \\
    Canes Venatici II & 4.259e+11 &  103 &   91 & $17.6^{+0.4}_{-0.4}$    & \checkmark              & \checkmark              & -                       & $17.65^{+0.40}_{-0.40}$ & $17.23^{+0.84}_{-0.68}$ & -                       & -                       \\
    Carina            & 4.363e+11 &  203 &  159 & $17.9^{+0.1}_{-0.1}$    & \checkmark              & \checkmark              & -                       & $17.99^{+0.34}_{-0.34}$ & $17.98^{+0.46}_{-0.28}$ & -                       & -                       \\
    Cetus II          & 3.737e+11 &   87 &   95 & $19.1^{+0.6}_{-0.6}$    & -                       & -                       & \checkmark              & -                       & -                       & -                       & -                       \\
    Columba I         & 4.024e+11 &  123 &  120 & $17.6^{+0.6}_{-0.6}$    & -                       & \checkmark              & -                       & -                       & -                       & -                       & -                       \\
    Coma Berenices    & 4.046e+11 &  115 &  151 & $19.0^{+0.4}_{-0.4}$    & \checkmark              & \checkmark              & -                       & $18.67^{+0.33}_{-0.32}$ & $18.52^{+0.94}_{-0.74}$ & $18.70^{+0.72}_{-0.69}$ & $21.59^{+0.26}_{-0.29}$ \\
    Draco             & 5.366e+11 &  175 &  150 & $18.8^{+0.1}_{-0.1}$    & \checkmark              & \checkmark              & -                       & $18.86^{+0.24}_{-0.24}$ & $19.09^{+0.39}_{-0.36}$ & $18.74^{+0.17}_{-0.16}$ & $21.52^{+0.26}_{-0.29}$ \\
    Draco II          & 5.607e+11 &  152 &  156 & $19.3^{+0.6}_{-0.6}$    & \checkmark              & \checkmark              & \checkmark              & -                       & $15.54^{+3.10}_{-4.07}$ & $18.87^{+0.17}_{-0.15}$ & -                       \\
    Eridanus II       & 4.173e+11 &   97 &   72 & $17.1^{+0.6}_{-0.6}$    & -                       & \checkmark              & \checkmark              & -                       & -                       & -                       & -                       \\
    Eridanus III      & 4.290e+11 &  107 &  113 & $18.1^{+0.6}_{-0.6}$    & -                       & -                       & \checkmark              & -                       & -                       & -                       & -                       \\
    Fornax            & 3.993e+11 &   92 &  125 & $17.8^{+0.1}_{-0.1}$    & \checkmark              & \checkmark              & \checkmark              & $18.15^{+0.16}_{-0.16}$ & $17.90^{+0.28}_{-0.16}$ & -                       & -                       \\
    Grus I            & 4.191e+11 &  109 &  105 & $17.9^{+0.6}_{-0.6}$    & -                       & \checkmark              & -                       & $17.96^{+0.90}_{-1.93}$ & -                       & -                       & -                       \\
    Grus II           & 4.203e+11 &  145 &  154 & $18.7^{+0.6}_{-0.6}$    & -                       & \checkmark              & -                       & -                       & -                       & -                       & -                       \\
    Hercules          & 4.330e+11 &  234 &  222 & $16.9^{+0.7}_{-0.7}$    & \checkmark              & \checkmark              & \checkmark              & $16.83^{+0.45}_{-0.45}$ & $16.28^{+0.66}_{-0.57}$ & -                       & -                       \\
    Horologium I      & 4.394e+11 &  110 &  132 & $18.2^{+0.6}_{-0.6}$    & \checkmark              & \checkmark              & -                       & $18.64^{+0.95}_{-0.39}$ & -                       & -                       & -                       \\
    Horologium II     & 4.272e+11 &  102 &  102 & $18.3^{+0.6}_{-0.6}$    & -                       & \checkmark              & -                       & -                       & -                       & -                       & -                       \\
    Hydra II          & 4.012e+11 &  205 &  162 & $17.8^{+0.6}_{-0.6}$    & \checkmark              & \checkmark              & \checkmark              & $16.56^{+0.87}_{-1.85}$ & $13.26^{+2.12}_{-2.31}$ & -                       & -                       \\
    Indus II          & 4.376e+11 &  216 &  257 & $17.4^{+0.6}_{-0.6}$    & -                       & \checkmark              & \checkmark              & -                       & -                       & -                       & -                       \\
    Kim 2             & 4.409e+11 &  198 &  201 & $18.1^{+0.6}_{-0.6}$    & -                       & -                       & \checkmark              & -                       & -                       & -                       & -                       \\
    Leo I             & 3.879e+11 &  128 &  138 & $17.8^{+0.2}_{-0.2}$    & \checkmark              & \checkmark              & \checkmark              & $17.80^{+0.28}_{-0.28}$ & $17.45^{+0.43}_{-0.23}$ & -                       & -                       \\
    Leo II            & 3.996e+11 &  111 &   83 & $18.0^{+0.2}_{-0.2}$    & \checkmark              & \checkmark              & \checkmark              & $17.44^{+0.25}_{-0.25}$ & $17.51^{+0.34}_{-0.28}$ & -                       & -                       \\
    Leo IV            & 3.670e+11 &  131 &  133 & $16.3^{+1.4}_{-1.4}$    & \checkmark              & \checkmark              & -                       & $16.64^{+0.90}_{-0.90}$ & $15.31^{+1.58}_{-2.90}$ & -                       & -                       \\
    Leo T             & 3.993e+11 &  130 &  122 & -                       & -                       & -                       & -                       & $17.32^{+0.38}_{-0.37}$ & $16.75^{+0.61}_{-0.53}$ & -                       & -                       \\
    Leo V             & 3.682e+11 &  130 &  145 & $16.4^{+0.9}_{-0.9}$    & \checkmark              & \checkmark              & \checkmark              & $16.94^{+1.05}_{-0.72}$ & $16.24^{+1.26}_{-1.36}$ & -                       & -                       \\
    Pegasus III       & 3.753e+11 &  160 &  168 & $17.5^{+0.6}_{-0.6}$    & -                       & \checkmark              & \checkmark              & -                       & -                       & -                       & -                       \\
    Phoenix II        & 4.314e+11 &  107 &   92 & $18.1^{+0.6}_{-0.6}$    & -                       & \checkmark              & \checkmark              & -                       & -                       & -                       & -                       \\
    Pictor I          & 4.344e+11 &  112 &  109 & $17.9^{+0.6}_{-0.6}$    & -                       & \checkmark              & \checkmark              & -                       & -                       & -                       & -                       \\
    Pisces II         & 3.718e+11 &  152 &  137 & $17.6^{+0.6}_{-0.6}$    & \checkmark              & \checkmark              & \checkmark              & $17.90^{+1.14}_{-0.80}$ & $15.94^{+1.25}_{-1.28}$ & -                       & -                       \\
    Reticulum II      & 4.423e+11 &  108 &  128 & $18.9^{+0.6}_{-0.6}$    & \checkmark              & \checkmark              & \checkmark              & $18.71^{+0.84}_{-0.32}$ & $17.76^{+0.93}_{-0.90}$ & -                       & $21.67^{+0.33}_{-0.30}$ \\
    Reticulum III     & 4.612e+11 &  125 &  158 & $18.2^{+0.6}_{-0.6}$    & -                       & \checkmark              & \checkmark              & -                       & -                       & -                       & -                       \\
    Sagittarius II    & 4.270e+11 &  319 &  312 & $18.4^{+0.6}_{-0.6}$    & -                       & \checkmark              & \checkmark              & -                       & -                       & -                       & -                       \\
    Sculptor          & 3.897e+11 &   88 &  114 & $18.5^{+0.1}_{-0.1}$    & \checkmark              & \checkmark              & -                       & $18.65^{+0.29}_{-0.29}$ & $18.42^{+0.35}_{-0.17}$ & -                       & -                       \\
    Segue 1           & 3.947e+11 &  128 &  154 & $19.4^{+0.3}_{-0.3}$    & \checkmark              & \checkmark              & \checkmark              & $19.41^{+0.39}_{-0.40}$ & $17.95^{+0.90}_{-0.98}$ & $19.81^{+0.93}_{-0.74}$ & $22.25^{+0.37}_{-0.62}$ \\
    Segue 2           & 4.072e+11 &  210 &  246 & -                       & -                       & -                       & -                       & $17.11^{+0.85}_{-1.76}$ & $13.09^{+1.85}_{-2.62}$ & -                       & -                       \\
    Sextans           & 3.699e+11 &  131 &  139 & $17.5^{+0.2}_{-0.2}$    & \checkmark              & \checkmark              & -                       & $17.87^{+0.29}_{-0.29}$ & $17.71^{+0.39}_{-0.21}$ & -                       & -                       \\
    Triangulum II     & 4.383e+11 &  187 &  198 & $19.1^{+0.6}_{-0.6}$    & \checkmark              & \checkmark              & -                       & -                       & $20.44^{+1.20}_{-1.17}$ & -                       & -                       \\
    Tucana II         & 4.518e+11 &  121 &  128 & $18.6^{+0.6}_{-0.6}$    & \checkmark              & \checkmark              & -                       & $19.05^{+0.87}_{-0.58}$ & -                       & -                       & -                       \\
    Tucana III        & 4.500e+11 &  110 &  132 & $19.3^{+0.6}_{-0.6}$    & -                       & \checkmark              & \checkmark              & -                       & -                       & -                       & -                       \\
    Tucana IV         & 4.517e+11 &  112 &  111 & $18.7^{+0.6}_{-0.6}$    & -                       & \checkmark              & \checkmark              & -                       & -                       & -                       & -                       \\
    Tucana V          & 4.593e+11 &  118 &  101 & $18.6^{+0.6}_{-0.6}$    & -                       & -                       & \checkmark              & -                       & -                       & -                       & -                       \\
    Ursa Major I      & 4.823e+11 &  110 &  108 & $17.9^{+0.5}_{-0.5}$    & \checkmark              & \checkmark              & -                       & $18.48^{+0.25}_{-0.25}$ & $17.48^{+0.42}_{-0.30}$ & $18.67^{+1.75}_{-1.02}$ & -                       \\
    Ursa Major II     & 5.594e+11 &  182 &  225 & $19.4^{+0.4}_{-0.4}$    & \checkmark              & \checkmark              & -                       & $19.38^{+0.39}_{-0.39}$ & $19.56^{+1.19}_{-1.25}$ & $19.50^{+0.29}_{-0.30}$ & -                       \\
    Ursa Minor        & 5.701e+11 &  146 &  123 & $18.9^{+0.2}_{-0.2}$    & \checkmark              & \checkmark              & -                       & $19.15^{+0.25}_{-0.24}$ & -                       & $19.12^{+0.15}_{-0.12}$ & $21.69^{+0.27}_{-0.34}$ \\
    Willman 1         & 4.771e+11 &  108 &  113 & $18.9^{+0.6}_{-0.6}$    & \checkmark              & \checkmark              & \checkmark              & $19.29^{+0.91}_{-0.62}$ & -                       & -                       & -                       \\
    \hline
  \end{tabular}
  \caption{Properties of each dSph.
    The columns give the name of the dSph, the average Fermi-LAT exposure, the average number of expected background events, the number of observed events in the dSph region, and the $J$-factors used in the various sets described in the text.
    \textit{Set 1a}, \textit{Set 1b}, and \textit{Set 1c} (labeled simply as ``a'', ``b'', and ``c'') are subsets of \textit{Set 1}, so we do not rewrite the value of the $J$-factor; instead, we indicate whether or not this dSph is included in the subset by a check mark.}
  \label{tab:targets}
\end{table}



\end{document}